Interferometric Measurement of Far Infrared Plasmons via Resonant Homodyne Mixing

G. C. Dyer[a], G. R. Aizin[b], S. J. Allen[c], A. D. Grine[a], D. Bethke[a], J. L. Reno[a], and E. A. Shaner[a]

[a]*Sandia National Laboratories, P.O. Box 5800, Albuquerque, New Mexico 87185*
[b]*Kingsborough College, The City University of New York, Brooklyn, New York 11235*
[c]*Institute for Terahertz Science and Technology, UC Santa Barbara, Santa Barbara, California 93106*

We present an electrically tunable terahertz two dimensional plasmonic interferometer with an integrated detection element that down converts the terahertz fields to a DC signal. The integrated detector utilizes a resonant plasmonic homodyne mixing mechanism that measures the component of the plasma waves in-phase with an excitation field functioning as the local oscillator. Plasmonic interferometers with two independently tuned paths are studied. These devices demonstrate a means for developing a spectrometer-on-a-chip where the tuning of *electrical length* plays a role analogous to that of physical path length in macroscopic Fourier transform interferometers.

1.  **Introduction**

There has been significant recent interest in the development of terahertz (THz) integrated circuits (ICs) and detectors based upon two dimensional electron gas (2DEG) systems in semiconductor nanostructures and graphene. Because microwave and THz fields coupled to a 2DEG excite plasma waves,[1-8] plasmon-based field effect devices can operate well above $f_T$, the cutoff frequency determined by carrier transit times.[9] Overdamped plasmonic field effect transistors (FETs) have been fabricated from III-V,[10-14] Si,[15, 16] and graphene[17] material systems and utilized for room temperature THz detection. To exploit underdamped two-dimensional (2D) plasmons in III-V heterostructures, cryogenic operation of a high electron mobility transistor (HEMT)[5, 7, 18-21] is generally required. Within this constraint, potential applications such as low-loss THz plasmonic ICs and resonant far infrared detectors based on III-V heterostructures can be realized in these material systems.

In this article, we demonstrate a two-dimensional (2D) plasmonic interferometer with an integrated resonant homodyne mixing element based upon a GaAs/AlGaAs HEMT with multiple gate terminals. Here it is demonstrated that biasing a gate in a HEMT near its threshold voltage while illuminated by THz radiation effectively produces a plasmonic homodyne mixing element and enables phase sensitive detection of plasma waves. When multiple plasmonic cavities are coupled to this gate-induced plasmonic mixing element, the device can be understood as a sub-wavelength two-path interferometer[7, 22] with an integrated on-chip detector where the paths are independently tuned. We also observe that unlike standard homodyne mixing techniques,[23] plasmonic homodyne mixing permits THz near field detection well above the conventional RC-limited bandwidth of the studied devices at their operational bias.

To describe the underlying mechanism of a solid state plasmonic interferometer, it is germane to first draw an analogy to an optical Mach-Zehnder interferometer. In Fig. 1(a), an optical Mach-Zehnder interferometer is diagrammed. Each optical path has a region of length $d$ where the permittivity $(\epsilon_D, \epsilon_S)$ and permeability $(\mu_D, \mu_S)$ of the electromagnetic medium is independently defined. If the phase velocity in these regions is given by $v_{D,S} = 1/\sqrt{\epsilon_{D,S}\mu_{D,S}}$, then the phase difference of beams on these two paths is $\theta_D - \theta_S = \omega d(v_D^{-1} - v_S^{-1})$. This phase difference results not from a difference in path lengths ($\Delta d = 0$), but instead from a difference in the electrical lengths of the paths ($\theta_D \neq \theta_S$). When the permittivity, permeability, or both, are tunable, then so are the electrical lengths $\theta_D$ and $\theta_S$ of these regions.

A transmission line circuit representing a pair of 2D plasmonic cavities coupled to a plasmonic mixing element is shown in Fig. 1(b). This representation of a 2D plasmonic HEMT is analogous to the optical Mach-Zehnder interferometer in Fig. 1(a) provided that the plasmonic cavities are driven in phase with equal amplitude and the variable inductances and resistances in each cavity are independently tunable. In both Figs. 1(a) and (b) a local oscillator (LO) field is coupled to a mixer to produce a down converted, or rectified, direct current (DC) signal by mixing with the fields incident from Path D and Path S. This homodyne mixing response containing two signal paths that are effectively 180 degrees out of phase can be understood through analyzing the response of a plasmonic HEMT as follows.

## 2. Resonant Plasmonic Homodyne Mixing in HEMTs

A resonant plasmonic photoresponse in the GaAs/AlGaAs HEMT pictured in Figs. 1 (c)-(e) under THz illumination may arise from several mechanisms. Recent studies have revealed a bolometric THz response mechanism,[24-26] while a photovoltage may also result from THz excitation when the 2DEG is at or near depletion.[21, 27] Here we focus upon this latter mechanism, a resonant plasmonic homodyne mixing photoresponse. The time-averaged mixing signal under THz illumination can be described in terms of the in-plane plasmonic voltages coupled to a region of 2DEG,[15, 28-30]

$$\delta V_{DS} = -G_{DS} \frac{\partial R_{DS}}{\partial V_{Gj}} \langle \delta V_{LO}(t) \, [\delta V_D(t) - \delta V_S(t)] \rangle. \tag{1}$$

The conductance $G_{DS}$ and resistance $R_{DS} = 1/G_{DS}$ between drain and source can be found from DC transport measurements. Near the threshold voltage of the $j^{th}$ gate $Gj$ the transport properties of the channel below $Gj$ become dominant relative to contributions from the contacts and the remainder of the channel. In this limit, $G_{DS}$ and $R_{DS}$ can then be taken to describe the transport in the plasmonic mixing region. The time dependent voltages in Eqn. 1 represent the THz fields coupled from opposing edges to the mixing region below $Gj$. The LO voltage $\delta V_{LO}(t)$ is capacitively coupled from $Gj$ to the 2DEG, while $\delta V_D(t) - \delta V_S(t)$ is the difference of the THz near fields coupled to the drain and source sides of the depleted region below gate $Gj$.

With $G2$ tuned to deplete the 2DEG below it as illustrated in Fig. 1(c), rectification takes place both at the left edge of $G2$ where the signal from Path D couples to the mixing region as well as at the right edge of $G2$ where the signal from Path S couples to the mixing region. Thus, the DC potential $\delta V_{DS}$ arises due to the difference between the rectified voltages on the drain-side of $G2$ and the source-side of $G2$. One of the underlying assumptions in this model is a loss of coherence between the two plasmonic signal channels when the mixing region of the 2DEG between them is biased near depletion. Recent modeling by Davoyan and Popov[31, 32] indicates that as $n_{2DEG} \to 0$, the plasmon near field amplitude decays rapidly from the edges of this region into its center. This isolates the plasma excitations at opposing edges from one another, and is consistent with the experimental assumption that these decoupled plasmonic fields produce independent mixing signals.

While in Fig. 1(c) $G2$ defines the mixing region of the device as annotated, in fact any of the gates $Gj$ can induce a plasmonic mixing region. In Figs. 1(d) and (e), alternative possibilities where $G1$ and $G3$, respectively, induce the mixing region are illustrated. The three possible choices for plasmonic mixing region of this device are explored through a combination of transport and photoresponse measurements at 8 K in Figs. 2 and 3. Measurements of the device transport were performed using a Stanford Research 830 lockin amplifier (LIA) to source 4.0 mV at 75.0 Hz to a 5.1 kOhm load resistor in series with the sample maintained at 8 K in a cryostat. The voltage drop across the load resistor was measured using the LIA to determine the device conductance as the sample gate biases were tuned. In Fig. 2(a), the device conductance as the voltage applied to $G1$ of the three-gate HEMT in Figs. 1(c)-(e) is tuned is shown. Though this two-point measurement includes contact resistances as well as series contributions from wire bonds and the external circuit, several key features directly related to the HEMT channel are evident. There is a discontinuity in the conductance near $V_{G1}$ = -0.95 V that indicates the presence of a parallel conduction channel in the device. In fact, the studied GaAs/AlGaAs heterostructure (Sandia wafer EA1149) has two quantum wells with a combined 2D electron density of 4.0 x $10^{11}$ cm$^{-2}$ that conduct in parallel. This feature results from the depletion of the quantum well nearest to the gate. The full depletion of both quantum wells below $G1$ is evident around $V_{G1}$ = -2.60 V. In this regime, transport in the region immediately below $G1$ dominates the system and contact resistances are negligible in comparison.

The DC measurement of the device conductance is connected with the expected THz photoresponse through Eq. 1. The factor $-G_{DS}\frac{\partial R_{DS}}{\partial V_{Gj}}$ in Eq. 1 relates the DC transport of a HEMT to its plasmonic mixing response, and is plotted in Fig. 2(b) as calculated from the conductance in Fig. 2(a). To verify that Eq. 1 and its corresponding transport measurement in Fig. 2(b) accurately describes the plasmonic mixing response, the photoresponse plotted in Fig. 2(c) was measured at 8 K with a 0.270 THz signal quasi-optically steered and focused on the device. The responsivity is calculated using the THz power incident on the window of the cryostat to normalize the measured voltage signal. A broadband THz antenna and a Si lens,[33] neither of which are shown, improve the coupling efficiency. Here the LIA modulated a continuous wave Virginia Diodes, Inc. Schottky diode multiplier millimeter wave source at 196.7 Hz and also measured the photovoltage generated between the source and drain terminals of the device under 0.270 THz illumination. Circuit loading effects due to the HEMT $RC$ time constant under typical bias conditions[34] become significant around several kHz modulation of the mm-wave source and reduce the measured photoresponse. Thus the conventional circuit $RC$ limited bandwidth is on the order of kHz, yet underdamped plasma excitations nonetheless provide coupling of THz fields to a high-resistance mixing region.

Through comparison of Figs. 2(b) and (c), it is evident that the measured photovoltage correlates strongly with the calculated transport curve, $-G_{DS}\frac{\partial R_{DS}}{\partial V_{G1}}$. Both data sets have maxima where the upper and lower quantum well channels below $G1$ are depleted, and also demonstrate an approximately three order of magnitude dynamic range. Although the plasmonic mixing response shown in Fig. 2 is largely unsurprising in light of the many demonstrations of this mechanism in highly varied transistor designs, material systems, and temperature ranges,[10, 11, 20, 29, 35-40] definitively establishing the origin of this photoresponse provides the basis for describing the operation of a two-path plasmonic interferometer. Asymmetry in the plasmonic signals coupled to the mixing region, $\delta V_D(t) - \delta V_S(t) \neq 0$, is required for generating a non-zero photovoltage.[41, 42] One means to explore introducing asymmetry into the device is by systematically voltage biasing each of the three gates.

In Fig. 3, the transport and responsivity characteristics at 8 K of the GaAs/AlGaAs HEMT in Figs. 1(c)-(e) are compared as one of the three gates is tuned independently while the other two are fixed at ground potential. The transport curves corresponding to Eq. 1 that are plotted in Fig. 3(a) are all nearly identical, consistent with the HEMT channel being homogeneous across the device and all three gates sharing an identical 2 μm width. Thus, the differences in the 0.270 THz responsivity shown in Fig. 3(b) arise due to asymmetry in the device induced via the applied gate bias. The responsivity with either gate $G1$ or gate $G3$, respectively, tuned is nearly identical in amplitude, but opposite in polarity. Taking $G1$ to define the mixing region as illustrated in Fig. 1(d), there are two plasmonic paths feeding into this mixing region: a path formed between S and $G1$ and a path formed between D and $G1$. Because these paths have different lengths, 2 μm vs. 10 μm, the phase and amplitude of monochromatic plasma waves impinging on the mixing region below $G1$ from opposing sides is, in general, non-identical. This produces a net photoresponse because $\delta V_D(t) - \delta V_S(t) \neq 0$. The scenario is similar when $G3$ defines the mixing region as shown in Fig. 1(e), but now the short and long plasmonic paths have exchanged relative positions in comparison to the first example. Consistent with the measured data, this inverts the signal polarity but leaves its amplitude largely unaffected.

A third possibility, pictured in Fig. 1(c), utilizes gate $G2$ to define the mixing region. In this case, the device is essentially symmetric about gate $G2$, though fabrication imperfections or misalignment of the incident radiation can introduce asymmetries. Here the photoresponse should be relatively weaker since the phase and amplitude of monochromatic plasma waves impinging on the mixing region from both paths will be nearly identical such that $\delta V_D(t) - \delta V_S(t) \approx 0$. In Fig. 3(b) the photoresponse with $G2$ tuned has a smaller amplitude, though its measureable amplitude indicates some asymmetry in the system under THz irradiation. Nonetheless, this is the most near to balanced configuration and also offers independent tunability of both Path D and Path S. Using this configuration, we can now systematically explore the operation of a monolithically integrated, balanced two-path plasmonic interferometer.

### 3. Two-Path Plasmonic Interferograms

With $G2$ biased to deplete the 2DEG below it, the plasmonic paths between S and $G2$ (Path S) and D and $G2$ (Path D) may be described in terms of tunable electrical lengths. Each of these paths is 6 μm long, with 2 μm regions below gates $G1$ and $G3$ that are voltage tuned. It is these sections below gates $G1$ and $G3$ that are of greatest interest, and it is useful to first relate applied gate voltages to 2DEG densities. Assuming a parallel plate capacitance between each gate and the 2DEG,

$$n_{1,3} = n_0 \frac{V_{th} - V_{G1,G3}}{V_{th}}, \qquad (2)$$

where $n_0$ is the intrinsic 2DEG density of 4.0 x $10^{11}$ cm$^{-2}$ and $V_{th}$ is the threshold voltage where the 2DEG is depleted, $V_{th} \cong -2.60\ V$. Since the equivalent distributed circuit elements in Fig. 1(c) depend directly upon $n_{1,3}$, the complex-valued transmission line propagation constants for Path S and Path D,

$$q_{S,D} = -i\sqrt{i\omega C_{1,3}(i\omega L_{1,3} + R_{1,3})}, \qquad (3)$$

may be defined for the voltage-tuned regions below $G1$ and $G3$, respectively. Here $L_{1,3} = m^*/e^2 n_{1,3}$, $R_{1,3} = L_{1,3}/\tau$ and $C_{1,3} = \epsilon q_{S,D}(1 + \coth q_{S,D} d)$ where $m^*$ is the electron effective mass of $0.067 m_e$, $e$ is the electron charge, $\tau$ is the plasmon damping time, $\epsilon$ is the permittivity of GaAs, and $d$ is the separation between the gates and the 2DEG.[43] In general, Eq. 3 is a transcendental equation, though it is identical to the standard definition of the propagation constant in transmission line theory as written in terms of equivalent circuit parameters. Though the total electrical lengths of Path S and Path D will also include the 4 μm of untuned 2DEG, it is sufficient to consider only the tuned regions to find the difference in electrical lengths. Thus, the relevant electrical lengths for Path S and D are,

$$\theta_{S,D} = a\, q'_{S,D}, \tag{4}$$

where $q_{S,D} = q'_{S,D} + iq''_{S,D}$, $q'_{S,D}$ and $q''_{S,D}$ are real, and $a = 2$ μm. Then the difference in electrical lengths of the two paths is $\Delta\theta = a\,[q'_D - q'_S]$. Physically, as either $G1$ or $G3$ is tuned towards threshold voltage, the electron density is decreased, the 2DEG (kinetic) inductance increases, the plasmon wavelength decreases, and the propagation constant increases.

The interferometric plasmonic signal with Path S and Path D independently controlled is illustrated in Figs. 4(a) and (b). The experimental measurements were performed at 8 K for excitation frequencies of 0.270 and 0.330 THz, respectively, with $V_{G2} = -2.55$ V. Because the gate voltage, the 2DEG density, the plasmon propagation constant, and the electrical length are all directly related by Eqs. 2-4, any of these may effectively parameterize the tuning of Paths S and D. In Fig. 4, the electrical lengths $\theta_{S,D}$ corresponding to Path S and Path D, respectively, are used in plotting the plasmonic interferogram. The diagonal lines from the lower left to upper right corners of Figs. 4(a) and (b) indicate where $\Delta\theta = 0$. For a balanced two-path interferometer, this diagonal marks where the signal should vanish as well as the boundary about which the signal should have anti-mirror symmetry. This is equivalent to stating that the signal $S(\theta_S, \theta_D)$ obeys the relation $S(\theta_1, \theta_2) = -S(\theta_2, \theta_1)$. Though the experimental results plotted in false color in Figs. 4(a) and (b) do not precisely follow this rule as would be the case in ideal balancing of the two paths, the qualitative picture nonetheless is indicative of anti-mirror symmetry. The signal tends to weaken then change polarity along $\Delta\theta = 0$ and for each positive signal polarity resonance at a coordinate $(\theta_1, \theta_2)$ there tends to be a companion resonance at $(\theta_2, \theta_1)$ with negative polarity.

The quantity $Re[\delta V_D - \delta V_S]$ calculated using a plasmonic transmission line model[43] as shown in Fig. 1(b) is plotted in Figs. 4(c) and (d) for excitation frequencies of 0.270 and 0.330 THz. Here it is assumed the antenna functions as a lumped element voltage source with an internal impedance found from its radiation resistance, $R_{RAD} \cong 72\,\Omega$.[26] Additionally, because the equivalent circuit sources driving the LO, Path S and Path D are in-phase, we take the real parts of the calculated plasmonic transmission line voltages to emulate the anticipated plasmonic mixing response. There is very good agreement between Figs. 4(a) and (c) using this approach, with several resonances matched in polarity observable along both the vertical and horizontal axes. However, the model interferogram in Fig. 4(d) does not correlate with Fig. 4(b) nearly as well. Although the lower order resonances seen at the shortest electrical lengths have the same polarity in Figs. 4(b) and (d), the model calculations predict additional higher order resonances that are not observed experimentally. Part of the discrepancy may arise from higher experimental plasmonic damping rates than the damping rate corresponding to an electron mobility of 100,000 cm$^{-2}$/V-s used in the model calculations. As the electrical length is increased by gate tuning, the losses increase, the resonances broaden, and the signal amplitude decreases. This is a qualitative feature of all plots in Fig. 4, and it is possible that the higher order modes in Fig. 4(d) cannot be resolved.

To further validate this proof-of-principle demonstration, we consider a second device design, shown in Fig. 5(a), where the two plasmonic paths are independently tunable four-period plasmonic crystals.[27] Here $G2$ is a single 2 µm gate, and $G1$ and $G3$ tune Path S and Path D, respectively, using four identical 2 µm wide gate stripes separated by 2 µm each. In this device, the gate tuning of plasma wave propagation cannot be interpreted as a simple changes of electrical length. Because plasmons are Bragg scattered in this short periodic lattice, a crystal quasi-momentum defined by the Bloch wavevector better describes plasma wave dispersion than the propagation constant of a plasmon below $G1$ or $G3$. The experimentally measured plasmonic interferogram in the left frame of Fig. 5(b) with 0.345 THz excitation and $G2$ biased to $V_{G2} = -2.80$ V at 8 K is therefore plotted in terms of gate voltages $V_{G1}$ and $V_{G3}$ to avoid ambiguity. This plasmonic interferometer can be understood as an in-situ plasmonic spectrometer for a more complicated plasmonic heterostructure than the device considered in Figs. 1(c)-(e). As before, plasmonic homodyne mixing takes place at the left and right edges of $G2$, but multi-period structures between S and $G2$ and D and $G2$ control the signals coupled to this mixing region. Despite the additional complexity of Paths S and D, a striking anti-mirror symmetry about $V_{G1} = V_{G3}$ where $S(V_1, V_2) = -S(V_2, V_1)$ is observed, indicating a well-balanced two-path plasmonic system.

A model interferogram of the calculated quantity $Re[\delta V_D - \delta V_S]$ is plotted in the right frame of Fig. 5(b) for an excitation frequency of 0.345 THz using a plasmonic transmission line model[43] to describe the four-period plasmonic crystals in Path S and Path D. Here a 2DEG density of 4.5 x $10^{11}$ cm$^{-2}$, about 10% larger than the 2DEG density determined from Hall measurements, and an electron mobility of 600,000 cm$^{-2}$/V-s, consistent with the mobility found from Hall measurements, were used in the model calculation. Although the overall agreement with experiment is largely qualitative in nature, the expected anti-mirror symmetry about $V_{G1} = V_{G3}$ where $S(V_1, V_2) = -S(V_2, V_1)$ is present. The most significant discrepancies between the model and experiment in Fig. 5(b) likely arise as a result of approximating the THz excitation as a lumped source in the transmission line model rather than a more realistic distributed excitation. While the transmission line approach predicts the resonant modes of the system with adequate fidelity, the exact plasmonic field amplitudes of Path S and D at the edges adjacent to the mixer will depend non-trivially upon the THz excitation of each plasmonic crystal. The THz coupling impacts not only the amplitudes of resonances, but also linewidths since radiative damping is a significant broadening mechanism. A lumped excitation is a reasonable approximation for plasmonic cavities with only several plasmonic elements, but limits the validity of the transmission line approach for modeling the plasmonic near fields of more complicated devices.

Several additional qualitative features in Fig. 5(b) prompt further consideration. First, in comparison to Fig. 4, many additional modes are observed with only a slight increase in excitation frequency. This is understood in part by comparing the 6 μm plasmonic path lengths in Fig. 1(c) to the 18 μm path lengths Fig. 5(a). The fundamental mode of the 18 μm path occurs at a lower frequency than that of the 6 μm path, and therefore a relatively denser set of higher order modes is anticipated for a given excitation frequency. Alternately, the coupling of four gated regions of the 2DEG in the device shown in Fig. 5(a) lifts a four-fold degeneracy, and therefore approximately four modes are expected for every one observed in the sample of Fig. 1(c). Additionally, the highest intensity signal is observed with significant tuning of gate voltage. This is analogous to observing the largest signal in Fig. 4 at any electrical lengths but the smallest measured. One possibility consistent with a recent study of localized modes in terahertz plasmonic crystals[27] is that specific modes in the spectrum couple less well to the mixing region as well as to the THz excitation field due to their confinement adjacent to contact

S or D. Although the distributed nature of the THz excitation precludes validation of this hypothesis using a lumped source to model the plasmonic near field amplitude, the non-monotonic behavior of signal intensities is suggestive of the localization of plasmon modes in Path S and Path D.

## 4. Conclusions

We have demonstrated an approach to integrate on-chip plasmonic interferometry with a widely-used plasmonic detection technique in this article. Although an antenna provided the distributed excitation of the signal channels and the LO of the plasmonic mixer, the presented approach should readily transfer to waveguide-coupled structures [7, 22] if the LO and signal channels are suitably isolated. The phase relationship between the LO and signal channels is determined by the coupling of the antenna excitation to HEMT terminals. Isolation of these channels would allow for control of their relative phase and potentially a quadrature measurement to extract both the amplitude and phase of an incident THz signal. This possibility arises because the plasmonic mixer is a field rather than power detector. While intensity interferograms are often measured by bringing two paths coincident upon a power detector, here field phase information is partially preserved by independently generating a DC signal from each path and reading out to a single differential channel.

The sensitivity of 2D plasma excitations to their environment portends intriguing possibilities for sensor development. Though the presented devices based on GaAs/AlGaAs heterostructures need both cryogenic cooling and a vacuum environment to operate, other plasmonic materials such as graphene have neither as a requirement. The electromagnetic screening of 2D plasma waves by a metal terminal is a limiting case of environment modifying plasmon dispersion. However, more subtle effects, particularly in graphene, can arise due to plasmon-phonon coupling with an adjacent material[44, 45] or the coupling of plasmons with an adsorbed polymer[46]. The research in this area thus far has focused on optical techniques. The devices presented in this article, however, suggest that an electro-optical approach to graphene near-field plasmonic sensing might also be fruitful.

Integration of interferometric elements into a voltage-tunable microelectronic plasmonic device provides potential advantages over a Fourier transform interferometer, particularly in the far infrared. Though the substantial reduction in optical path length is beneficial, the most significant advantage is provided by the broad voltage tunability. Optical interferometers must mechanically tune a path over lengths on the order of meters to measure high resolution spectra. In the studied low dimensional plasmonic interferometers, the electrical length can comfortably tune over an order of magnitude, though the intrinsic plasmonic losses ultimately limit resolution. Despite this limitation, this study provides an important step towards device and system integration of multiple signal path plasmonic detectors for future generations of far infrared sensing technologies.

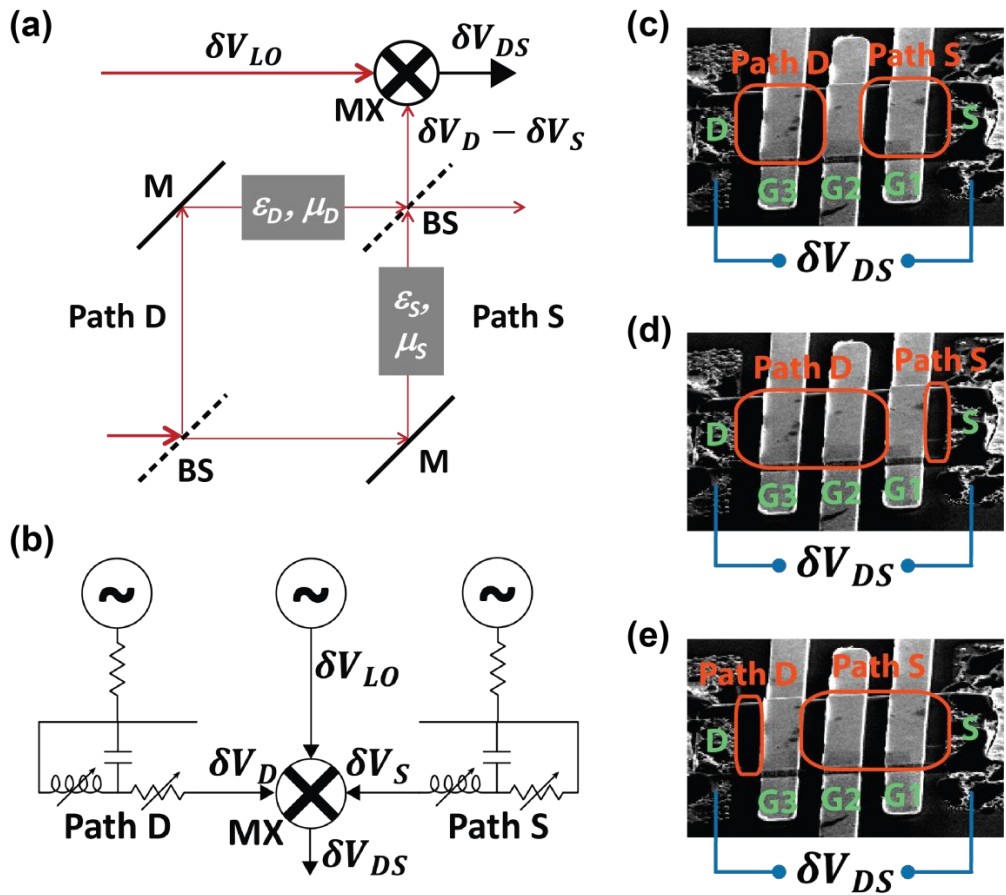

**Figure 1**. (a) Layout of an optical Mach-Zehnder interferometer where the electromagnetic properties of Path D and Path S are independently defined. Beam splitters are labeled BS and mirrors are labeled M. Readout in this schematic is accomplished using a mixer, labelled MX, to produce a DC signal at the second beamsplitter where the recombined beams are 180 degrees out of phase. (b) An equivalent circuit schematic for a two-path plasmonic interferometer where Path D and Path S are independently tunable. Path S, Path D and the mixer's local oscillator are all excited in phase, and a DC signal is produced by the mixer MX. (c) A scanning electron micrograph of a two-path plasmonic interferometer where gate G2 of a HEMT defines the mixing element and Path S and Path D are tuned by G1 and G3, respectively. The gates are all approximately 2 μm wide and separated by 2 μm. The distance between the Ohmic contacts S and D is 14 μm. In (d) and (e) the same device is shown but with G1 and G3, respectively, defining the mixing region.

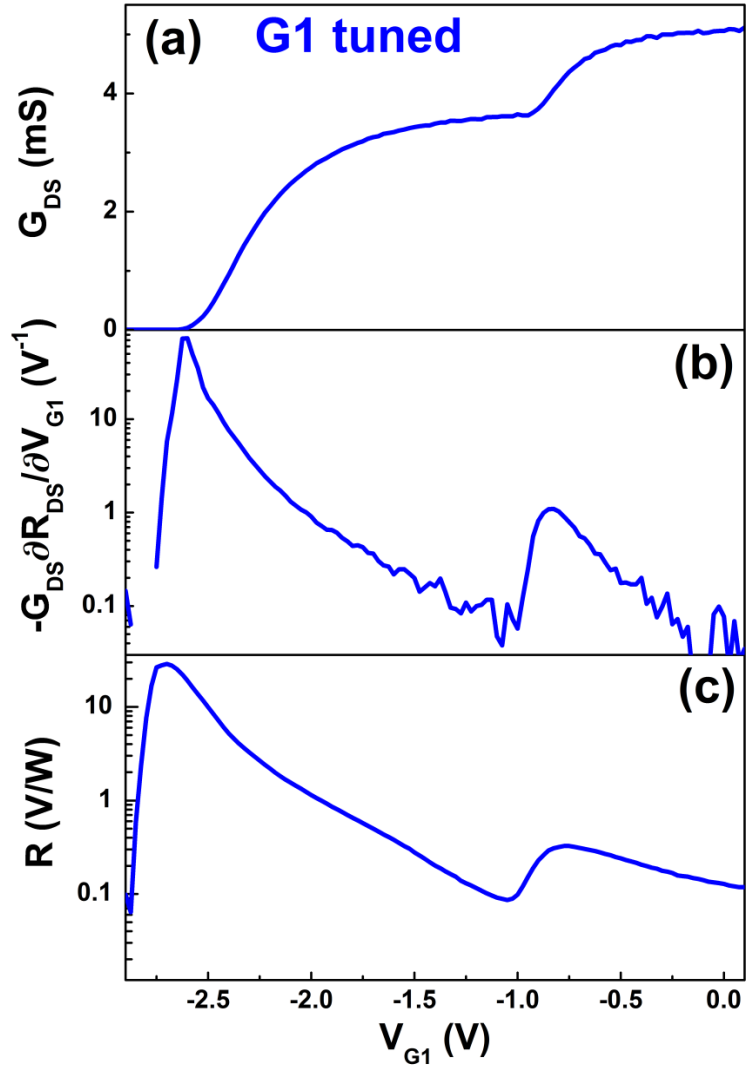

**Figure 2**. (a) The channel conductance at 8 K of the HEMT illustrated in Fig. 1 is plotted as a function of voltage applied to gate G1. (b) Using the channel conductance measured at 8 K, the product $-G_{DS}\frac{\partial R_{DS}}{\partial V_{G1}}$ is calculated and plotted as a function of voltage applied to G1. (c) The 8 K device photoresponse under 0.270 THz illumination is plotted as a function of voltage applied to G1.

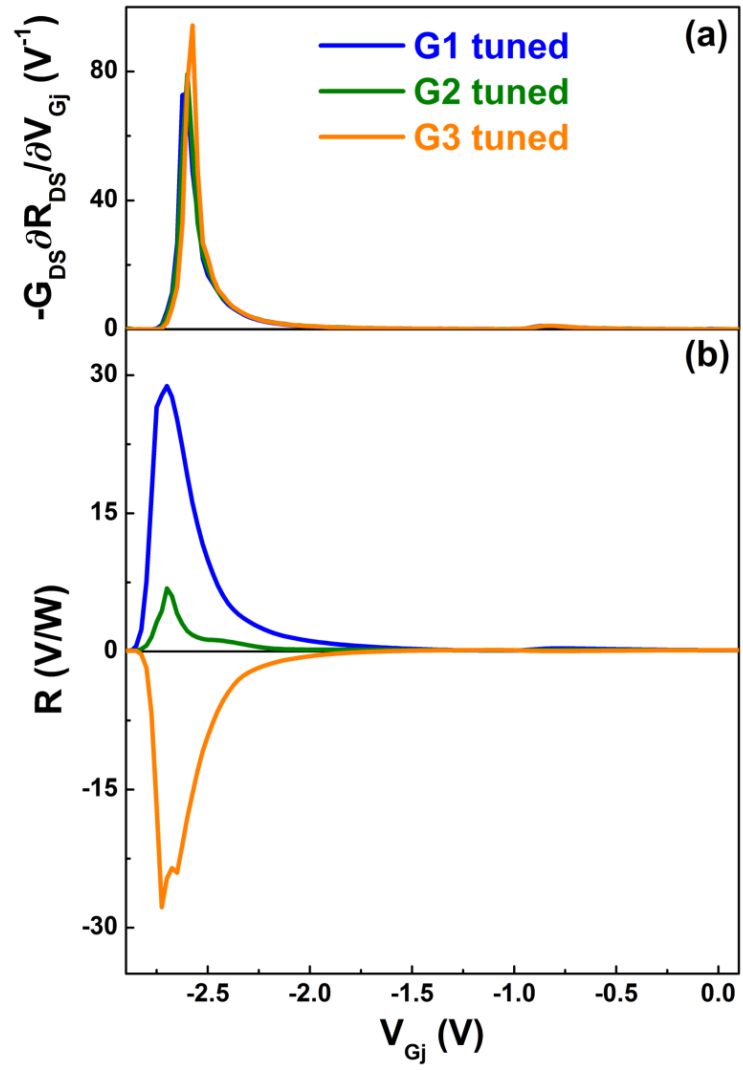

**Figure 3**. (a) Using the channel conductance measured at 8 K of the HEMT illustrated in Fig. 1, the product $-G_{DS}\frac{\partial R_{DS}}{\partial V_{Gj}}$ is calculated and plotted as a function of voltage applied to gates G1, G2 and G3. (b) The 8 K device photoresponse under 0.270 THz illumination is plotted as a function of voltage applied to gates G1, G2 and G3. For all measurements, one gate was tuned whiled the other two were held at ground potential.

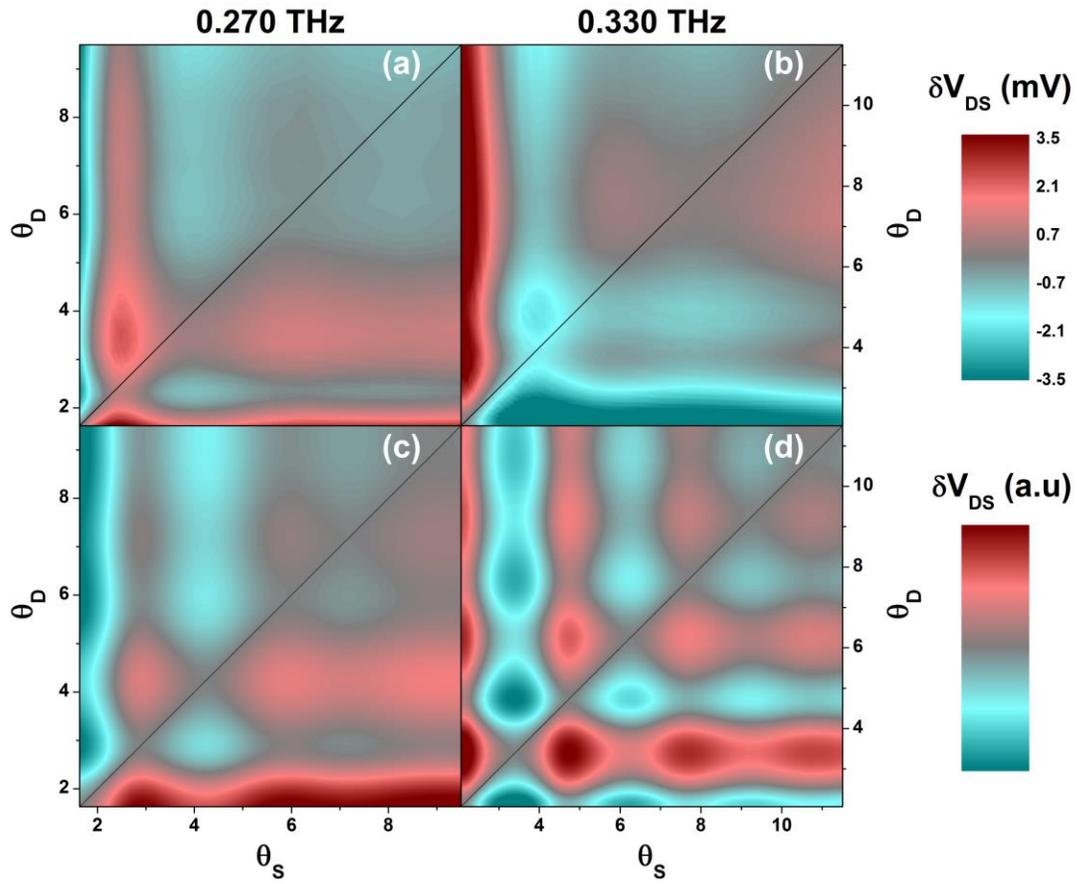

**Figure 4**. With G2 defining the mixing region of the HEMT as illustrated in Fig. 1(c), the photoresponse under (a) 0.270 THz and (b) 0.330 THz excitation at 8 K is mapped as a function of the electrical lengths of Path S, $\theta_S$, and Path D, $\theta_D$. A model calculation of the photoresponse under (c) 0.270 THz and (d) 0.330 THz is also plotted using a transmission line formalism to describe the independent signals from Paths S and D coupled to the mixing element below G2.

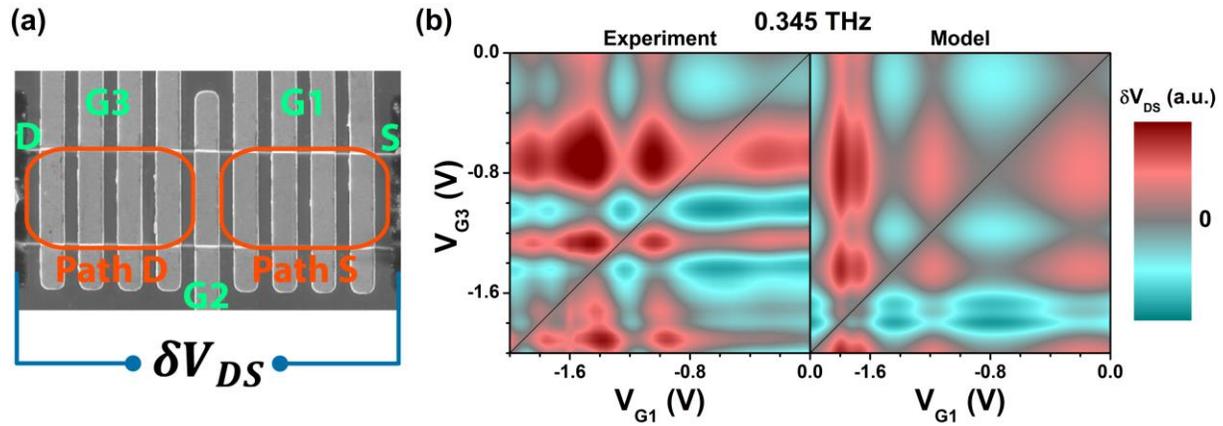

**Figure 5**. (a) A scanning electron micrograph of a two-path plasmonic crystal interferometer where gate G2 of the HEMT defines the mixing element and Path S and Path D are tuned by G1 and G3, respectively. Both G1 and G3 consist of four identically tuned gate stripes that are 2 μm wide and separated by 2 μm. The distance between the Ohmic contacts S and D is 34 μm. (b) The 8 K device photoresponse under 0.345 THz illumination is mapped as a function of voltage applied to gates G1 and G3 with G2 fixed at -2.80 V in the left frame. A model calculation of the photoresponse under 0.345 THz illumination is also plotted in the right frame using a transmission line formalism to describe the independent signals from Paths S and D coupled to the mixing element below G2.